\documentclass{IEEEtran}

\usepackage{hyperref}
\usepackage{amsmath}
\usepackage{cancel}
\usepackage{tikz}
\usepackage{circuitikz}
\usepackage[utf8]{inputenc}
\usepackage{color}
\usepackage{lipsum}
\usepackage{soul}
\usepackage{cite}
\usepackage{diagbox}

\newcommand{\C}[1]{\mathbf{\hat{#1}}}
\newcommand{\M}[1]{\mathbf{#1}}
\newcommand{\T}[1]{\mathrm{#1}}
\newcommand{\V}[1]{\boldsymbol{#1}}

\newcommand{\AV}[1]{\bar{\bar{\V{#1}}}}
\renewcommand{\u}[1]{\boldsymbol{\hat{#1}}}
\newcommand{\BS}{\Psi}

\title{Conversion Matrix Method of Moments for Time-Varying Electromagnetic Analysis}
\author{S. F.~Bass,~\IEEEmembership{Student Member,~IEEE,}
        A. M.~Palmer,~
        K. R.~Schab,~\IEEEmembership{Member,~IEEE,}
        K. C.~Kerby-Patel,~\IEEEmembership{Senior Member,~IEEE,}
        and~J. E.~Ruyle,~\IEEEmembership{Senior Member,~IEEE}
\thanks{Manuscript received  \today; revised \today.}
\thanks{S. F.~Bass and J. E.~Ruyle are with and A. M.~Palmer was formerly with the School of Electrical and Computer Engineering and the Advanced Radar Research Center (ARRC), University of Oklahoma, Norman, OK, USA (e-mail: ruyle@ou.edu and sfbass@ou.edu).}
\thanks{K. R. Schab is with the Department of Electrical and Computer Engineering, Santa Clara University, Santa Clara, CA, USA (e-mail: kschab@scu.edu).}
\thanks{K. C. Kerby-Patel is with the Engineering Department, University of Massachusetts Boston, Boston, MA USA (e-mail: kc.kerby-patel@umb.edu) }

}

\begin{document}


\maketitle
\setcounter{page}{1}

\begin{abstract}
    A conversion matrix approach to solving network problems involving time-varying circuit components is applied to the method of moments for electromagnetic scattering analysis.  Detailed formulations of this technique's application to the scattering analysis of structures loaded with time-varying circuit networks or constructed from general time-varying media are presented.  The computational cost of the method is discussed, along with an analysis of compression techniques capable of significantly reducing computational cost for partially loaded systems.  Several numerical examples demonstrate the capabilities of the technique along with its validation against conventional methods of modeling time-varying electromagnetic systems, such as finite difference time domain and transient circuit co-simulation.
\end{abstract}


\section{Introduction}
\IEEEPARstart{T}{ime-varying} or nonlinear (non-LTI) electromagnetic structures are capable of exhibiting unique behaviors beyond the capabilities of their linear time-invariant (LTI) counterparts.  Examples of non-LTI antenna systems include the use of active and non-Foster matching \cite{bahr1977use, nordholt1980new, zhu2012}, direct antenna modulation \cite{daly2018tuning, schab2019pulse, santos2020}, and 
time-varying loading whose modulation rate is comparable to the carrier or antenna resonant frequency (as opposed to the symbol rate) \cite{hopf1981fast, wang2007, loghmannia2019parametric, slevin2020, singletary2021}. 
The aforementioned methods all involve locally time-varying or nonlinear loading, though the effects of distributed time-variation, i.e., space-time modulated materials, have also been explored \cite{koutserimpas2018electromagnetic,ramaccia2018nonreciprocity,caloz2019spacetime_pt1,caloz2019spacetime_pt2,shcherbakov2019,chamanara2019,wu2019serrodyne,taravati2020space}.

Most modeling of time-varying electromagnetic structures relies on time-domain techniques such as the finite difference time domain method (FDTD), transient circuit co-simulation, and time-domain method of moments \cite{landt1983time}. While these techniques are accurate and extremely general, they have certain disadvantages that motivate the development of alternative modeling strategies \cite{du2019simulation}. For example, full-wave transient analyses have few opportunities for partial simulation re-use between variations of time-varying properties. Additionally, these methods do not directly represent frequency domain phenomena frequently employed in the design of LTI systems (e.g., steady state radiated power, network parameters), though these can be obtained via Fourier transformations. By contrast, frequency-domain techniques often afford significant opportunities for partial simulation re-use and their formulations naturally align with many common frequency domain metrics. In particular, the direct connection between the method of moments (MoM) and the dyadic Green's function makes it favorable in applications such as modal current analysis \cite{garbacz1971generalized,harrington1971theory}, automated design synthesis \cite{rahmat1999electromagnetic,ethier2014antenna,capek2019shape}, and the development of fundamental bounds on LTI system performance \cite{gustafsson2016antenna,gustafsson2019tradeoff}. Connecting these benefits to a method compatible with the analysis of linear, time-varying systems may greatly accelerate the study of non-LTI electromagnetic structures in similar application areas.

 Our approach to meeting this need is to hybridize MoM with existing frequency-domain methods developed for time-varying circuit analysis. 
 Specifically, we use conversion matrices to transform a MoM-based scattering problem into one involving an LTI $N$-port network interfaced with time-varying subcircuits.  Previous work in hybridizing MoM with conversion matrix solvers exists, but is relatively limited to special cases focused on sparse lumped loading. In \cite{huang1993analysis,jayathurathnage2021timevarying}, well-known time-varying and non-linear circuit analysis techniques (conversion matrices and harmonic balance) were used to model a single load at the feed point of an antenna. By collapsing the antenna to a lumped impedance, this method is extremely efficient, though it constrains loading to a single location. In \cite{epp1992}, a periodic structure with periodically time-varying loads is treated by a similar conversion matrix / MoM (CMMoM) hybridization but again with a focus on sparse lumped loading. While these methods are useful and reflect a common practical implementation of time-varying systems through local time-varying elements, they do not directly allow for generalization to systems involving multiple loads or distributed time-varying properties.  Modeling of time-varying material properties has been studied in an analytical context equivalent to CMMoM, but this treatment is limited to the analysis of conducting cylinders \cite{salary2018time}.
 
Based in part on preliminary studies in \cite{palmer2019investigation}, here, we develop a generalized CMMoM method allowing for multiple lumped and distributed spatiotemporal loading of electromagnetic structures of arbitrary shape.  A core component of this method is the use of conversion matrices, a well known technique in time-varying circuit analysis.  Because this technique is less common in electromagnetics communities, we review its fundamentals in Sec.~\ref{sec:cm} and establish notation used throughout the paper. In Sec.~\ref{sec:mom}, we describe the integration of conversion matrices with MoM for lumped and distributed loading, followed by a discussion of the source-frequency and harmonic-frequency power quantities in Sec.~\ref{sec:power}.  In Sec.~\ref{sec:computational} we discuss issues of computational cost and compression and in Sec.~\ref{sec:examples}, we present three examples to illustrate the range of problems that may be analyzed by this technique.  We conclude in Sec.~\ref{sec:conclusion} with discussion of potential applications, limitations, and extensions of the presented method.

\section{Conversion matrix methods}
\label{sec:cm}
Conversion matrices enable frequency-domain modeling of systems with time-varying components by describing the coupling between voltages and currents at multiple frequencies~\cite{maas2003nonlinear}. Their use in circuit design is well documented, but these techniques are rarely applied to electromagnetic scattering problems. Here we review the fundamentals of conversion matrix methods on multiport networks 
in preparation for their application to open, distributed electromagnetic systems via MoM.

\subsection{Lumped time-varying elements}
\label{sec:cm-single}

When a time-varying voltage is impressed across a time-varying load, the spectral content of the resulting current corresponds to a mixing of the applied voltage with the time-variation of the load. This can be seen by applying the convolution theorem to Ohm's law, as in
\begin{equation}
    i(t) = v(t) g(t)
\end{equation}
and
\begin{equation}
    I(\omega) = \int_{-\infty}^{\infty} V(\omega-\omega')G(\omega')\T{d}\omega'
    \label{eq:tvconv}
\end{equation}
where $i(t)$, $v(t)$, and $g(t)$ are the time-domain current, voltage, and conductance of the load, and $I(\omega)$, $V(\omega)$, and $G(\omega)$ are their Fourier transforms. In an LTI system, the conductance has only a static component $G(\omega)\sim\delta(\omega)$, and the current can only contain frequencies that are present in the voltage excitation. When the conductance $g(t)$ is not static, the resulting current includes sum and difference mixing products of the voltage and load frequency content. 

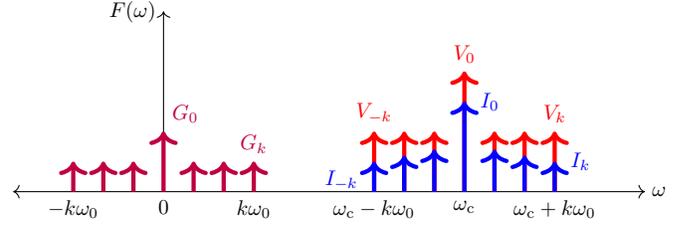
\begin{figure}
    \centering
        \begin{tikzpicture}[scale=0.8,transform shape]
    \draw[<->] (-2.5,0)--(8,0) node[right]{$\omega$};
    \draw[->] (0,0)--(0,3) node[left]{$F(\omega)$};
    \draw[->,purple,ultra thick] (0,0) node[below,black]{$0$} -- (0,1) node[above right]{$G_0$};
    \draw[->,purple,ultra thick] (0.5,0) -- (0.5,0.5);
    \draw[->,purple,ultra thick] (1,0) -- (1,0.5);
    \draw[->,purple,ultra thick] (1.5,0) node[below,black]{$k\omega_0$} -- (1.5,0.5) node[above] {$G_k$};
    \draw[->,purple,ultra thick] (-0.5,0) -- (-0.5,0.5);
    \draw[->,purple,ultra thick] (-1,0) -- (-1,0.5);
    \draw[->,purple,ultra thick] (-1.5,0) node[below,black]{$-k\omega_0$} -- (-1.5,0.5) ;
    
    \draw[->,red,ultra thick] (5,0) node[below,black]{$\omega_\T{c}$} -- (5,2) node[above] {$V_0$};
    
    \draw[->,blue,ultra thick] (5,0) -- (5,1.5) node[right] {~$I_0$};
    
    \draw[red, ->, ultra thick] (5.5,0) -- (5.5,1);
    \draw[red, ->, ultra thick] (6,0) -- (6,1);
    \draw[red, ->, ultra thick] (6.5,0) node[below,black] {$\omega_\T{c} + k\omega_0$} -- (6.5,1) node[above,red] {$V_{k}$};
    
        \draw[red, ->, ultra thick] (4.5,0) -- (4.5,1);
    \draw[red, ->, ultra thick] (4,0) -- (4,1);
    \draw[red, ->, ultra thick] (3.5,0) node[below,black] {$\omega_\T{c} - k\omega_0$} -- (3.5,1) node[above,red] {$V_{-k}$};
    
    \draw[blue, ->, ultra thick] (5.5,0) -- (5.5,0.7);
    \draw[blue, ->, ultra thick] (6,0) -- (6,0.6);
    \draw[blue, ->, ultra thick] (6.5,0)  -- (6.5,0.5) node[right,blue] {~$I_{k}$};    
    
    \draw[blue, ->, ultra thick] (4.5,0) -- (4.5,0.7);
    \draw[blue, ->, ultra thick] (4,0) -- (4,0.6);
    \draw[blue, ->, ultra thick] (3.5,0)  -- (3.5,0.5) node[below left,blue] {$I_{-k}~$};
    
    \end{tikzpicture}
    
    \caption{Frequency convention used throughout this paper.  Time-varying loads are represented in Fourier series of the fundamental frequency $\omega_0$, e.g., $\{G_k\}$, while current and voltage indexing centers around a modulation frequency $\omega_\T{c}$.}
    \label{fig:FConv}
\end{figure}

The preceding discussion is valid for loads with arbitrary time dependence. If the load's time variation is periodic, it may be represented by a Fourier series, as in
\begin{equation}
    g(t) = \sum_{k = -K}^{K} G_k \T{e}^{\T{j}k\omega_0 t}
\end{equation}
and
\begin{equation}
    G(\omega) = \sum_{k=-K}^K G_k \delta(\omega-k\omega_0),
    \label{eq:contf}
\end{equation}
where $g(t)$ is the time-varying conductance of the load, $G_k$ is the $k$th Fourier coefficient, $\omega_0$ is the fundamental frequency of the time-varying component, and $K$ is large enough to contain sufficient frequency-domain content. Similarly, we may expand the voltage in terms of a series of $\omega_0$ harmonics centered about a reference frequency $\omega_\T{c}$, 
\begin{equation}
    v(t) = \sum_{k = -K}^{K} V_k \T{e}^{\T{j}(\omega_\T{c}+k\omega_0) t}
\end{equation}
and
\begin{equation}
    V(\omega) = \sum_{k=-K}^K V_k \delta(\omega-\omega_\T{c}-k\omega_0)
    \label{eq:voltf}
\end{equation}
so long as the baseband representation of the driving voltage is periodic in the fundamental frequency $\omega_0$. If the voltage is not periodic in $\omega_0$, then the excitation can be decomposed into multiple problems with different center frequencies. In this paper, we focus on single frequency excitation, where this condition is naturally satisfied as $V_k = 0$ for all $k\neq 0$. As a consequence of centering the harmonics about a reference frequency $\omega_c$, as in~\eqref{eq:voltf}, the negative frequency components of the excitation signal are ignored. Instead we focus on the upper sideband as shown in Fig.~\ref{fig:FConv}. If desired, contributions from negative frequencies may be calculated by a secondary calculation~\cite{maas2003nonlinear}. 

Adopting the same expansion and notation for the current $i(t)$, the conductance relationship in~\eqref{eq:tvconv} may be written as
\begin{equation}
    I_k = \sum_{\ell =-L}^{L} V_{k-\ell } G_\ell
    \label{eq:cm-single-eq}
\end{equation}
and in matrix form as
\begin{equation}
    \begin{bmatrix}
        I_{-K} \\
        I_{1-K} \\
        \vdots \\
        I_{K}
    \end{bmatrix}
    =
    \begin{bmatrix}
        G_0 & G_{-1} & \hdots & G_{-2K}\\
        G_1 & G_0 & \hdots &  G_{1-2K} \\
        \vdots & \vdots & \ddots & \vdots \\
        G_{2K} & G_{2K-1} & \hdots & G_0
    \end{bmatrix}
    \begin{bmatrix}
        V_{-K} \\
        V_{1-K} \\
        \vdots \\
        V_{K}
    \end{bmatrix}
    \label{eq:CMcon}
\end{equation}
or more compactly
\begin{equation}
    \C{I} = \C{G}\C{V}
\end{equation}
where $\C{G}$ is the conversion matrix representation of the time-varying conductance $g(t)$.
This matrix models the modulating effect of the time-varying component, where the $k$th element of the current vector contains contributions from every $G_p V_q$ product that satisfies $p+q=k$. 

An expression similar to \eqref{eq:CMcon} can be derived using a time-varying resistance, rather than conductance~\cite{maas2003nonlinear}. This illustrates an inverse relationship between resistive and conductive conversion matrices, similar to that of their LTI counterparts,
\begin{equation}
    \C{R} = \C{G}^{-1}.
    \label{eq:CMResCon}
\end{equation}

Conversion matrices may also be generated for time-varying inductors and capacitors, with the general structure 
\begin{equation}
    \C{I} = \mathrm{j}\C{\Omega} \C{C}\C{V} \qquad \mathrm{and} \qquad \C{V} = \mathrm{j}\C{\Omega} \C{L}\C{I}
    \label{eq:CMCapInd}
\end{equation}
where
\begin{equation}
    \C{\Omega} = 
    \begin{bmatrix}
        \omega_{-K} & 0 & \hdots & 0 \\
        0 & \omega_{1-K} & \hdots & 0 \\
        \vdots & \vdots & \ddots & \vdots \\
        0 & 0 & \hdots &\omega_{K}
    \end{bmatrix}, \quad \omega_k = \omega_\T{c}+k\omega_0,
    \label{eq:cm-cap-and-ind}
\end{equation}
and the matrices $\C{C}$ and $\C{L}$ are capacitance and inductance conversion matrices of the form of the matrix $\C{G}$ in \eqref{eq:CMcon}. 
The conversion matrices of conductances $\C{G}$, resistances $\C{R}$, capacitances $\C{C}$, and inductances $\C{L}$, can be treated as basic lumped components and combined into larger networks by following usual series and parallel circuit rules~\cite{maas2003nonlinear}.  For real-valued time-varying circuit elements, the matrices $\C{G}$, $\C{R}$, $\C{L}$, and $\C{C}$ are naturally Hermitian symmetric. However, multiplication of $\C{C}$ or $\C{L}$ by the frequency matrix $\C{\Omega}$ or its inverse, as in \eqref{eq:CMCapInd}, breaks the Hermitian symmetry of the impedance conversion matrices of time-varying inductive or capacitive elements.

\subsection{Loaded multi-port networks}
\label{sec:cm-multiport}

The time-domain representation of an LTI, $N$-port network with time-varying resistors on each port may be written as
\begin{equation}
    v_\alpha(t) = i_\alpha(t) r_\alpha(t) + \sum_{\beta=1}^N z_{\alpha\beta}(t)\star i_\beta(t)
    \label{eq:MPtime}
\end{equation}
where $v_\alpha(t)$, $i_\alpha(t)$, and $r_\alpha(t)$ are the time-varying voltage, current, and resistance across port $\alpha$, and  $z_{\alpha\beta}(t)$ is the open-circuit impedance impulse response between ports $\alpha$ and $\beta$.  This translates to a frequency-domain representation
\begin{multline}
    V_\alpha(\omega) = \int_{-\infty}^\infty I_\alpha(\omega-\omega') R_\alpha(\omega')\T{d}\omega' \\+ \sum_{\beta=1}^N Z_{\alpha\beta}(\omega) I_\beta(\omega)
\end{multline}
where $V_\alpha(\omega)$, $I_\alpha(\omega)$, $R_\alpha(\omega)$, and $Z_{\alpha\beta}(\omega)$ are the frequency domain forms of the parameters in~\eqref{eq:MPtime}.

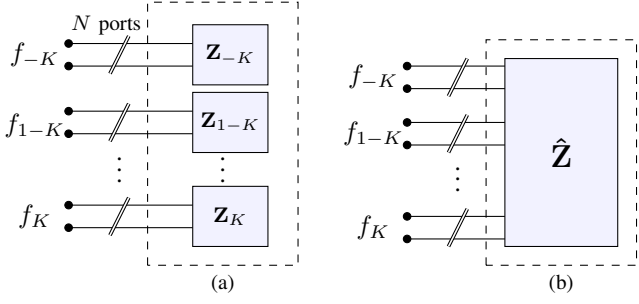
\begin{figure}
\centering
\begin{tikzpicture}

\begin{scope}[shift = {(4.5,3.75)}]
\draw (3, -0.25) node {\footnotesize (b)};
\draw[dashed] (2, 0) rectangle ++(2,3);
\draw[fill=blue!5] (2.25, 0.25) rectangle ++(1.5, 2.5);
\draw (2.25, 0.35) -- ++(-1.3, 0);
\draw (2.25, 0.65) -- ++(-1.3, 0);
\filldraw[color=black] (0.95, 0.35) circle (0.05);
\filldraw[color=black] (0.95, 0.65) circle (0.05);
\draw[double] (1.5, 0.25) -- ++(0.25, 0.5);
\draw (0.5, 0.5) node {$f_K$}; 

\draw (2.25, 2.35) -- ++(-1.3, 0);
\draw (2.25, 2.65) -- ++(-1.3, 0);
\filldraw[color=black] (0.95, 2.35) circle (0.05);
\filldraw[color=black] (0.95, 2.65) circle (0.05);
\draw[double] (1.5, 2.25) -- ++(0.25, 0.5);
\draw (0.5, 2.5) node {$f_{-K}$}; 
 
\draw (2.25, 1.6) -- ++(-1.3, 0);
\draw (2.25, 1.9) -- ++(-1.3, 0);
\filldraw[color=black] (0.95, 1.6) circle (0.05);
\filldraw[color=black] (0.95, 1.9) circle (0.05);
\draw[double] (1.5, 1.5) -- ++(0.25, 0.5);
\draw (0.5, 1.75) node {$f_{1-K}$}; 
\draw (3, 1.5) node {\large $\C{Z}$}; 
\draw (1.625, 1.25) node {$\vdots$};
\end{scope}

\draw (3, 3.5) node {\footnotesize (a)};
\draw[dashed] (2, 3.75) rectangle ++(2,3.5);
\draw[fill=blue!5] (2.6, 4.0) rectangle ++(1,0.8) node[pos=.5] {\footnotesize $\M{Z}_K$}; 
\draw (2.6, 4.25) -- ++(-1.7, 0);
\draw (2.6, 4.55) -- ++(-1.7, 0);
\filldraw[color=black] (0.95, 4.25) circle (0.05);
\filldraw[color=black] (0.95, 4.55) circle (0.05);
\draw[double] (1.5, 4.15) -- ++(0.25, 0.5);
\draw (0.5, 4.4) node {$f_K$}; 

\draw (3, 5.125) node {$\vdots$}; 
\draw (1.625, 5.125) node {$\vdots$};

\draw[fill=blue!5] (2.6, 6.15) rectangle ++(1,0.8) node[pos=.5] {\footnotesize $\M{Z}_{-K}$}; 
\draw (2.6, 6.4) -- ++(-1.7, 0);
\draw (2.6, 6.7) -- ++(-1.7, 0);
\filldraw[color=black] (0.95, 6.4) circle (0.05);
\filldraw[color=black] (0.95, 6.7) circle (0.05);
\draw[double] (1.5, 6.3) -- ++(0.25, 0.5);
\draw (0.5, 6.55) node {$f_{-K}$}; 

\draw[fill=blue!5] (2.6, 5.25) rectangle ++(1,0.8) node[pos=.5] {\footnotesize $\M{Z}_{1-K}$}; 
\draw (2.6, 5.5) -- ++(-1.7, 0);
\draw (2.6, 5.8) -- ++(-1.7, 0);
\filldraw[color=black] (0.95, 5.5) circle (0.05);
\filldraw[color=black] (0.95, 5.8) circle (0.05);
\draw[double] (1.5, 5.4) -- ++(0.25, 0.5);
\draw (0.5, 5.6) node {$f_{1-K}$}; 

\draw (1.45, 6.95) node {\footnotesize $N$ ports};
\end{tikzpicture}

    \caption{Diagram of LTI multiport system (a) and periodically time-varying multiport system (b) as represented by a conversion matrix. Double slash across port symbol denotes $N$ physical ports. Each physical port supports voltages and currents at each of the $2K+1$ harmonic frequencies.  An LTI system with $2K+1$ frequencies  can be represented by $2K+1$ independent linear systems, or a block diagonal conversion matrix, because there is no conversion between frequencies.}
    \label{fig:portsandfreqs}
\end{figure}

After manipulations closely resembling those in~\eqref{eq:contf} and~\eqref{eq:voltf}, we obtain the equation
\begin{multline}
    V_\alpha(\omega_c+k\omega_0) =
    \sum_{\ell =-L}^{L} I_\alpha(\omega_c+(k-\ell )\omega_0)R_\alpha(\ell \omega_0) \\
    + \sum_{\beta=1}^N Z_{\alpha\beta}(\omega_c+k\omega_0)I_\beta(\omega_c+k\omega_0),
\end{multline}
and after including the frequency notation from~\eqref{eq:CMcon} as superscripts, the previous expression may be rewritten as
\begin{equation}
    V_\alpha^k = \sum_{\ell =-L}^{L} I_\alpha^{k-\ell } R_\alpha^\ell + \sum_{\beta=1}^N Z_{\alpha\beta}^k I_\beta^k.
    \label{eq:cm-multiport-single-eq}
\end{equation}
Equations of this form can be collected into a matrix form by grouping the port voltages and currents at each frequency. The resulting system of equations reads
\begin{multline}
\begin{bmatrix}
    \M{V}^{-K}\\
    \M{V}^{1-K}\\
    \vdots\\
    \M{V}^K
\end{bmatrix}
= 
\begin{bmatrix}
    \M{R}^{0} & \M{R}^{-1} & \hdots & \M{R}^{-2K} \\ 
    \M{R}^{1} & \M{R}^{0} & \hdots & \M{R}^{1-2K}\\
    \vdots  & \vdots & \ddots & \vdots \\
    \M{R}^{2K} & \M{R}^{2K-1} &  \hdots & \M{R}^0  \\
\end{bmatrix}
\begin{bmatrix}
    \M{I}^{-K}\\
    \M{I}^{1-K}\\
    \vdots\\
    \M{I}^K
\end{bmatrix}\\
+
\begin{bmatrix}
    \M{Z}^{-K} & 0 & \hdots & 0  \\
    0 & \M{Z}^{1-K} & \hdots & 0  \\
    \vdots & \vdots & \ddots  & \vdots \\
    0 & 0 & \hdots & \M{Z}^{K} \\
\end{bmatrix}
\begin{bmatrix}
    \M{I}^{-K}\\
    \M{I}^{1-K}\\
    \vdots\\
    \M{I}^K
\end{bmatrix}
\label{eq:CMexpand}
\end{multline}
and has a similar structure to the conversion matrices of~\eqref{eq:CMcon}, with the key difference that every element within each matrix or vector is replaced by a submatrix or subvector of dimension $N$. The submatrices and subvectors take the forms
\begin{subequations}
\begin{equation}
    \M{R}^k =
    \begin{bmatrix}
        R_1^k & 0 & \hdots & 0 \\
        0 & R_2^k & \hdots & 0 \\
        \vdots & \vdots & \ddots & \vdots \\
        0 & 0 & \hdots & R_N^k 
    \end{bmatrix},
    \label{eq:CMsuperR}
\end{equation}
\begin{equation}
    \M{Z}^k = \left[Z_{\alpha\beta}^k \right],\quad
    \M{V}^k = \left[V_{\alpha}^k \right],\quad
    \M{I}^k = \left[I_{\alpha}^k \right].
    \label{eq:cm-components}
\end{equation}
\end{subequations}
Thus, $\M{R}^k$ is a diagonal matrix consisting of the $k$-th harmonic of the time-varying resistances at all $N$ ports, $\M{Z}^k$ is the open-circuit impedance matrix of the $N$-port LTI network at the $k$-th harmonic centered about $\omega_c$, and $\M{V}^k$ and $\M{I}^k$ contain voltages and currents existing on all $N$ ports at the $k$-th harmonic, as illustrated in Fig.~\ref{fig:portsandfreqs}(b). The system of equations in \eqref{eq:CMexpand} may be expressed in a more compact form as
\begin{equation}
    \C{V} = \left(\C{R}+\C{Z}\right)\C{I}.
    \label{eq:CMcompact}
\end{equation}
For the case of purely LTI loading, we have $\C{R}^{k\neq 0}=0$ and the matrices in \eqref{eq:CMexpand} reduce to a block diagonal matrix as shown in Fig.~\ref{fig:portsandfreqs}(a). As a result, the system is represented by $2K+1$ decoupled matrix equations at each harmonic.  While reciprocal LTI networks lead to symmetric conversion matrices $\C{Z}$, the conversion matrices for real-valued loads are Hermitian symmetric based on the conjugate symmetry of their Fourier representations.  Thus, unless loads are selected specifically to have real-valued Fourier spectra, the system conversion matrix $\C{Z}+\C{R}$ is neither symmetric nor Hermitian. Representations similar to~\eqref{eq:CMcompact} may be constructed for arbitrary networks of time-varying resistances, capacitances, and inductances using the forms in \eqref{eq:CMResCon} and~\eqref{eq:CMCapInd} along with standard circuit element combination rules \cite{maas2003nonlinear}.

If desired, the multi-port conversion matrix can be grouped by port rather than by frequency \cite{maas2003nonlinear}. This arrangement would lead to an overall matrix structure that resembled an open-circuit impedance matrix, with each element in the matrix replaced by a small conversion matrix, i.e.,
\begin{equation}
    \begin{bmatrix}
        \C{V}_{1} \\
        \C{V}_{2} \\
        \vdots \\
        \C{V}_{N}
    \end{bmatrix}
    =
    \begin{bmatrix}
        \C{Z}_{11} & \C{Z}_{12} & \hdots & \C{Z}_{1N} \\
        \C{Z}_{21} & \C{Z}_{22} & \hdots & \C{Z}_{2N} \\
        \vdots & \vdots & \ddots & \vdots \\
        \C{Z}_{N1} & \C{Z}_{N2} & \hdots & \C{Z}_{NN}
    \end{bmatrix}
    \begin{bmatrix}
        \C{I}_{1} \\
        \C{I}_{2} \\
        \vdots \\
        \C{I}_{N}
    \end{bmatrix}
    \label{eq:CMportwise}
\end{equation}
where $\C{V}_\alpha$, $\C{I}_\alpha$, and $\C{Z}_{\alpha\beta}$ are conversion matrix parameters as defined in~\eqref{eq:CMcon}, but specific to the $\alpha$ and $\beta$ ports of the $N$-port network. The matrices in~\eqref{eq:CMexpand} and~\eqref{eq:CMportwise} share the same elements, but are re-ordered to emphasize different relationships. While other work in multiport conversion matrices use a port-wise arrangement \cite{maas2003nonlinear,jiang2006conversion}, this work uses the format of~\eqref{eq:CMexpand} to facilitate compatibility with standard MoM techniques, as will be discussed in the next section.

\section{Method of Moments and conversion matrices}
\label{sec:mom}

\begin{figure}
    \centering
    \includegraphics[width=3.25in,clip,trim = 0.4in 1in 0.5in 0.25in]{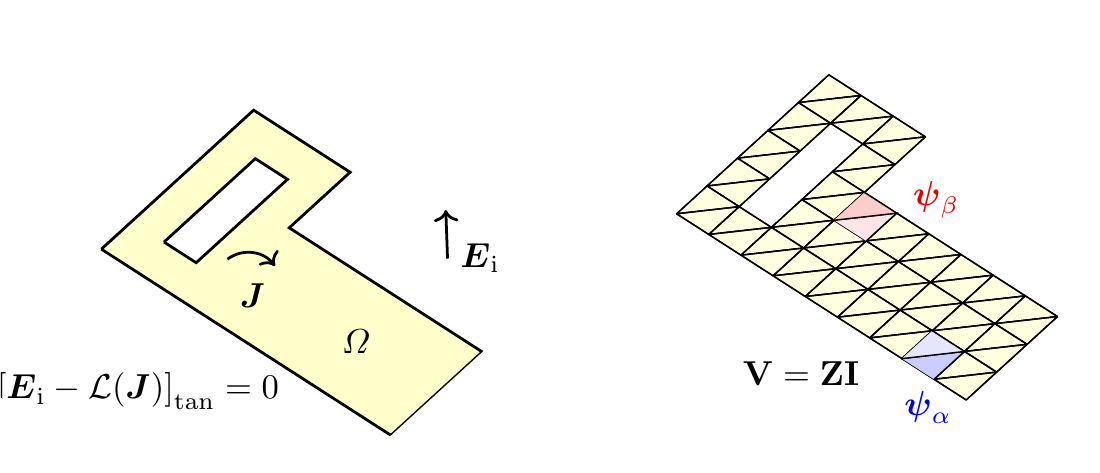}
    \caption{Discretization of a structure $\varOmega$ (left) into finite elements supporting localized basis functions $\{\V{\psi}_n\}$ (right).  RWG basis functions \cite{rao1982electromagnetic} are depicted, with each basis function centered on one mesh edge and spanning two adjacent triangles.}
    \label{fig:mom-schem}
\end{figure}

A broad class of LTI electromagnetic scattering problems may be recast as LTI network problems through the use of the method of moments (MoM) \cite{harrington1993field}. Here we consider problems involving a perfectly conducting (PEC) surface $\varOmega$ supporting surface currents $\V{J}$, as shown in the left panel of Fig.~\ref{fig:mom-schem}.  To solve for the surface currents induced by a monochromatic incident field $\V{E}_\T{i}$, we may expand the surface current into an appropriate basis $\{\V{\psi}_n\}$ in order to convert Maxwell's equations into a matrix form of the electric field integral equation
\begin{equation}
    \M{V}(\omega) = \M{Z}(\omega)\M{I}(\omega)
    \label{eq:mom-vzi}
\end{equation}
where $\M{V}$ and $\M{I}$ are vectors containing coefficients related to the incident field and induced current, respectively, $\omega$ is the excitation frequency, and $\M{Z}$ is the impedance matrix representing the scattered field operator $\mathcal{L}(\V{J})$~\cite{harrington1993field}.  Throughout this paper we assume Galerkin testing is applied such that the impedance matrix is transpose symmetric.  The frequency dependence of all quantities, to be dropped in all future expressions, explicitly describes the LTI nature of the scatterer and indicates that currents will only exist at the excitation frequency.  Induced currents due to multi-tone excitation can be analyzed by direct superposition of weighted monochromatic solutions, i.e., Fourier series or transforms.

\subsection{Loading}
\label{sec:mom-loading}

When the chosen basis is sufficiently localized, e.g., when pulse \cite{yeh1967theory}, or RWG basis functions \cite{rao1982electromagnetic} are used, the elements of the vectors $\M{V}$ and $\M{I}$ may be interpreted as voltages and currents present at discrete locations, or ports, on the structure, as shown in the right panel of Fig.~\ref{fig:mom-schem}.  For example, in the case of RWG basis functions, each port corresponds to one edge within the triangularized mesh created from the original structure $\varOmega$.

With the aforementioned network interpretation of the electromagnetic scattering problem in Fig.~\ref{fig:mom-schem}, lumped loading at any combination of the scatterer's ports is straightforward to model via the addition of a diagonal loading matrix to the impedance matrix $\M{Z}_\T{L}$, 
\begin{equation}
    \M{Z}\rightarrow \M{Z}+\M{Z}_\T{L},
\end{equation}
whose elements are related to the lumped element loading at each port~\cite{jiao1999fast}.  Loads of finite size spanning multiple basis functions may also be modeled through the use of non-diagonal loading matrices~\cite{lo2011finite}.  An identical approach also allows for the modeling of non-PEC surfaces, which may be inhomogeneous and/or anisotropic, characterized by surface impedance $Z_\T{s}$ through a non-diagonal loading matrix related to the Gram matrix of the chosen basis \cite{jiao1999fast, gustafsson2016antenna}. 

\subsection{Lumped time-varying loads}
\label{sec:mom-tv-lumped}

The techniques outlined in Secs.~\ref{sec:mom-loading} and \ref{sec:cm-multiport} may be combined to model time-varying lumped elements loading any or all ports of a MoM network representation of the structure $\varOmega$.  The resulting conversion matrix system takes on precisely the same form as \eqref{eq:CMexpand}--\eqref{eq:CMcompact}, where $\C{V}$ and $\C{I}$ are stacked vectors containing fields and currents at all harmonic frequencies and $\C{R}$ and $\C{Z}$ are the dynamic (associated with time-varying loading) and static (associated with the underlying LTI structure) conversion matrices, respectively.   

Here we again note that the static conversion matrix $\C{Z}$ is block diagonal and contains impedance matrices representing the structure $\varOmega$ at each harmonic frequency.  The individual blocks $\M{R}^k$ of the dynamic loading matrix $\C{R}$ are diagonal for localized bases, with off diagonal blocks $\M{R}^{k\neq 0}$ representing Fourier components of each element's time variation.  In the special cases when the structure is unloaded or if all loading elements are static, the system becomes LTI and the system conversion matrix becomes block diagonal, reducing to a set of $2K+1$ decoupled matrix equations, each involving only quantities at a single frequency.  Generalization to capacitive and inductive loads follows the form of \eqref{eq:cm-cap-and-ind}, and again combinations of LTI and/or time-varying components may be synthesized by standard circuit analysis rules \cite{maas2003nonlinear}.

\subsection{Distributed space-time loading}
\label{sec:mom-tv-dist}

Like the extension of lumped LTI loading concepts to the modeling of LTI distributed material parameters (e.g., surface impedances), conversion matrix methods can also be applied to model time-varying distributed material parameters within a MoM framework.  Following the examples discussed in Sec.~\ref{sec:cm}, we begin with a description of this approach for resistive material properties and then extend the method to reactive behavior at the end of this section.

Consider a structure $\varOmega$ constructed of a material with space-time varying anisotropic surface resistivity $\AV{r}_\T{s}(\V{r},t)$.  The time-domain surface current $\V{j}(\V{r},t)$ is determined by the total tangential field $\V{e}^\T{tan}(\V{r},t)$ via the boundary condition
\begin{equation}
\AV{r}_\T{s}(\V{r},t)\V{j}(\V{r},t) = \V{e}^\T{tan}(\V{r},t),\quad \V{r}\in\varOmega
\label{eq:mom-tv-surf-td}
\end{equation}
where the anisotropic resistivity tensor may be written as
\begin{equation}
\AV{r}_\T{s}(\V{r},t) = \begin{bmatrix}
r_\T{s}^{uu} & r_\T{s}^{uv} \\
r_\T{s}^{vu} & r_\T{s}^{vv} \\
\end{bmatrix}
\end{equation}
with $u$ and $v$ representing a two dimensional local coordinate system on the surface being considered.  Note that in the special case of zero surface resistivity this equation reduces to the electric field integral equation for PEC surfaces.  Separating the total tangential field into incident and scattered components, we may write \eqref{eq:mom-tv-surf-td} in the frequency domain as
\begin{multline}
    \u{n}\times\V{E}_\T{inc}(\V{r},\omega) = \\\u{n} \times \left[ \AV{R}_\T{s}(\V{r},\omega)\star\V{J}(\V{r},\omega) + \mathcal{L}_\omega (\V{J})(\V{r}) \right],\quad \V{r} \in \varOmega.
\label{eq:mom-tv-surf-fd}
\end{multline}
where the linear operator $\mathcal{L}_\omega$ returns the negative scattered field from the current distribution at frequency $\omega$~\cite{harrington1993field}, $\u{n}$ is a unit vector normal to the surface of $\varOmega$, and the tensor convolution is understood to represent four scalar convolutions according to standard matrix multiplication rules.

As carried out in previous sections, we assume that the temporal variation of the surface resistivity is representable via a Fourier series in the fundamental frequency $\omega_0$, i.e.
\begin{equation}
    \AV{R}_\T{s}(\V{r},\omega) = \sum_{k=-K}^K \AV{R}_{\T{s}}^k(\V{r}) \delta(\omega-k\omega_0).
\end{equation}
Substituting this representation into \eqref{eq:mom-tv-surf-fd}, expanding the surface current at all frequencies into the basis $\{\psi_\alpha\}$ and applying Galerkin testing, we obtain the linear system
\begin{equation}
V_{\alpha}^k = \sum_\beta \sum_{\ell=-L}^{L} R_{\T{s},\alpha\beta}^{\ell} I_{\beta}^{k-\ell} + \sum_\beta Z_{\alpha\beta}^k I_{\beta}^k
\end{equation}
where
\begin{equation}
R_{\T{s},\alpha\beta}^k = \int_V \V{\psi}_\alpha(\V{r}') \cdot \AV{R}_{\T{s}}^k(\V{r}') \cdot \V{\psi}_\beta(\V{r}')\, \T{d} V,
\label{eq:rs-element-anisotropic}
\end{equation}
and the quantities $V_\alpha^k$ and $Z^k_{\alpha\beta}$ are exactly those used to construct the incident field vector and impedance matrix within the MoM matrix equation \eqref{eq:mom-vzi} at frequency $\omega_k = \omega_\T{c}+k\omega_0$.  We recognize the above expression as a generalization of the multi-port network relation \eqref{eq:cm-multiport-single-eq}, where now the time-varying elements are allowed to relate currents and voltages across multiple ports.  Adapting the notation of \eqref{eq:CMexpand}, \eqref{eq:cm-components}, and \eqref{eq:CMcompact}, we arrive at the conversion matrix system 
\begin{equation}
\C{V} = \left(\C{R}_\T{s}+\C{Z}\right)\C{I},
\label{eq:rs-cm}
\end{equation}
where the key differentiating feature, as compared to the multi-port network in \eqref{eq:CMsuperR}, is that the block matrices $\M{R}_\T{s}^k$ comprising the dynamic conversion matrix $\C{R}_\T{s}$ are no longer strictly diagonal.  When the surface resistance is isotropic and spatially homogeneous, \eqref{eq:rs-element-anisotropic} becomes a prescription for a scaled form of the Gram matrix of the chosen basis $\{\V{\psi}_\alpha\}$, with the nature of off-diagonal terms dependent solely on the extent of non-orthogonality of the basis.

In the frequency domain, lossy reactive polarization is represented by a complex resistivity or complex susceptibility.  It is beyond the scope of this work to explore the dynamics and possibilities of how lossy reactive media might be made time-varying.  However, we can note that a simple damped oscillator model of a polarization process, i.e., the Lorentz-Drude model, gives rise to a boundary condition of the form
\begin{multline}
\alpha(\V{r},t)\frac{\partial}{\partial t }\V{j}(\V{r},t) + \beta(\V{r},t)\V{j}(\V{r},t) \\+ \kappa(\V{r},t)\int_{-\infty}^t\V{j}(\V{r},t')\T{d}t' = \V{e}^\T{tan}(\V{r},t)
\end{multline}
where $\alpha$, $\beta$, and $\kappa$ are space-time-varying material parameters.  Here we have opted to not use the standard physical parameters (e.g., damping constant, plasma frequency) of the Lorentz-Drude model since it is not known how individual parameters may be made time-varying and what the physical implications of those variations might be.  Applying the method of moments and conversion matrix techniques used in previous sections, we find that this system reduces to the form
\begin{equation}
\C{V} = \left(\T{j}\C{\Omega}\C{A}+\C{B}-\T{j}\C{K}\C{\Omega}^{-1} + \C{Z}\right)\C{I} .
\end{equation}
The above expression, unsurprisingly, resembles that of an $N$-port network loaded with time-varying series RLC oscillators; the primary difference being basis function overlap terms leading to non-diagonal matrices $\C{A}$, $\C{B}$, and $\C{K}$.  Any of the parameters $\alpha$, $\beta$, and $\kappa$ may be made anisotropic, leading to matrix elements of the form of \eqref{eq:rs-element-anisotropic}.

\section{Interpretation of power quantities} \label{sec:power}
Much like in the study of LTI antennas or circuit networks, many physically relevant quantities can be obtained through linear or quadratic forms of terminal currents and voltages using conversion matrix methods for time-varying networks.  Consider an LTI structure with conversion matrix $\C{Z}$ loaded with time-varying elements represented by the conversion matrix $\C{Z}_\T{tv}$, i.e.,
\begin{equation}
    \C{V} = \left(\C{Z}+\C{Z}_\T{tv}\right)\C{I}.
    \label{eq:power-vzi}
\end{equation}
Due to the orthogonality of sinusoids at dissimilar frequencies, cross-frequency voltage-current products do not contribute to time-average power flow. Thus, power quantities within the system have the same form as in standard LTI problems, and may be written as a sum of the individual harmonic powers $P^k$, e.g.,
\begin{equation}\label{eq:power1}
    P=\frac{1}{2}\T{Re}\{\C{I}^\T{H}\C{V}\} = \sum_{k=-K}^K P^k,
\end{equation}
where
\begin{equation}
    P^k = \frac{1}{2} \T{Re}\{\M{I}^{k,\T{H}}\M{V}^k\}.
\end{equation}
We may interpret the power $P$ as the total power removed from the incident field, i.e., extinction power or that supplied by the excitation field \cite{Jackson:100964, gustafsson2020upper}. For single-frequency sources, only the source frequency may contribute to this power, as all elements of $\C{V}$ are zero except for the source frequency term $\M{V}^0$, and therefore $P=P^0$.   

By \eqref{eq:power-vzi}, the power $P$ may also be written
\begin{equation}
    P=\frac{1}{2}\C{I}^\T{H}(\C{R}+\C{R}_\T{tv})\C{I},
\end{equation}
where $\C{R}$ and $\C{R}_\T{tv}$ are the Hermitian parts of $\C{Z}$ and $\C{Z}_\T{tv}$, respectively. By the block diagonal nature of the static conversion matrix $\C{Z}$, the total power dissipated in the LTI portion of the structure may be interpreted as a linear sum of powers dissipated by currents at each frequency, i.e.,
\begin{equation}
    P_\T{LTI}=\frac{1}{2}\C{I}^\T{H}\C{R}\C{I} = \sum_{k=-K}^K P_\T{LTI}^k
    \label{eq:plti}
\end{equation}
where
\begin{equation}
    P_\T{LTI}^k = \frac{1}{2}\M{I}^{k,\T{H}}\M{R}^k\M{I}^k.
\label{eq:pltik}
\end{equation}
Assuming the LTI portion of the system is passive, the matrix $\C{R}$ is positive semidefinite, as are its submatrices $\M{R}^k$ at every harmonic frequency, and $P_\T{LTI}^k\geq 0$ for all $k$.  Note that dissipation at each frequency in the LTI portion of this system can be decomposed into contributions from thermal losses (absorption) and radiation (scattering), each with a corresponding quadratic form similar to \eqref{eq:pltik}, see \cite{gustafsson2016antenna,jelinek2016optimal}.

Similarly, the power dissipated in the time-varying portion of the structure is
\begin{multline}
    P_\T{tv}=\frac{1}{2}\C{I}^\T{H}\C{R}_\T{tv}\C{I}\\ = \frac{1}{2} \sum_{k=-K}^K \M{I}^{k,\T{H}} \sum_{\ell=-K}^K  \M{R}_\T{tv}^{k-\ell}\M{I}^\ell = \sum_{k=-K}^K P_\T{tv}^k.
\end{multline}
In this case, the matrix $\C{R}_\T{tv}$ may be indefinite, and off-diagonal blocks prevent the writing of the power $P_\T{tv}$ as a sum of quadratic forms in each harmonic current $\M{I}^k$, as was possible for LTI dissipation in \eqref{eq:plti}-\eqref{eq:pltik}. In problems with single-frequency sources at the $k=0$ harmonic,  $P_\T{tv}^k = -P_\T{LTI}^k$ for $k \neq 0$ and $P_\T{tv}^0$ may be either positive or negative. Thus, the time-varying element must supply power at intermodulation frequencies, while at the source frequency, it may either accept or supply power \cite{manley1956}.

\section{Compression and computational cost}\label{sec:computational}

A typical MoM solution of an $N$-port network requires inversion of an $N$-dimensional matrix, resulting in a na\"ive\footnote{In practice, advanced algorithms and the structure of a matrix itself can often be exploited to accelerate inversion to $\mathcal{O}(N^\alpha)$, with $1 \leq \alpha \leq 3$.  For brevity, here we consider only the nominal worst-case scenario of $\alpha = 3$.} computational cost of $\mathcal{O}(N^3)$. The MoM-conversion matrix method proposed in this work increases the size of the matrix to be inverted by the number of calculated harmonics $N_\T{f} = 2K+1$, leading to a significantly increased inversion cost of the order $\mathcal{O}(N^3 N_\T{f}^3)$.

However, often the problem of interest involves an LTI system that is loaded with a small number of time-varying loads $N_\T{l} \ll N$. In this case, it is useful to compress the system into the smallest number of degrees of freedom possible before inverting the system matrix. 

\subsection{Compression Techniques}

We begin by partitioning the single frequency impedance matrix representing the LTI portion of the system as 
\begin{equation}
    \begin{bmatrix}
        \M{V}_\T{u} \\
        \M{V}_\T{l}
    \end{bmatrix}
    =
    \begin{bmatrix}
        \M{Z}_\T{uu} & \M{Z}_\T{ul}\\
        \M{Z}_\T{lu} & \M{Z}_\T{ll}
    \end{bmatrix}
    \begin{bmatrix}
        \M{I}_\T{u} \\
        \M{I}_\T{l}
    \end{bmatrix}
    \label{eq:CmpBeforeMat}
\end{equation}
where the subscripts $\T{l}$ and $\T{u}$ denote the ports to be loaded or left unloaded. Rearranging the top line of the above expression into the form
\begin{equation}
    \M{I}_\T{u} = \M{Z}_\T{uu}^{-1} \left(\M{V}_\T{u} - \M{Z}_\T{ul} \M{I}_\T{l} \right)
    \label{eq:CmpBeforeMatTop}
\end{equation}
shows that the current on the unloaded portions of the structure can be written in terms of only the excitation $\M{V}_\T{u}$ and loaded port currents $\M{I}_\T{l}$. Substituting~\eqref{eq:CmpBeforeMatTop} into the bottom line of~\eqref{eq:CmpBeforeMat} gives
\begin{equation}
    \left( \M{Z}_\T{ll}-\M{Z}_\T{lu}\M{Z}_\T{uu}^{-1}\M{Z}_\T{ul} \right)\M{I}_\T{l} = \M{V}_\T{l} - \M{Z}_\T{lu}\M{Z}_\T{uu}^{-1}\M{V}_\T{u} ,
    \label{eq:CmpBefore}
\end{equation}
or more compactly,
\begin{equation}
    \mathbf{\check{Z}}_\T{l}\mathbf{I}_\T{l} = \mathbf{\check{V}}_\T{l},
    \label{eq:CmpBeforeCpt}
\end{equation}
where $\check{~}$ represents a compressed quantity.  The system of equations in \eqref{eq:CmpBeforeCpt} serves as a compressed $N_\T{l}$-dimensional representation of the loaded portion of the system at a single frequency.  The ``hidden'' degrees of freedom associated with the unloaded ports may be easily recovered via \eqref{eq:CmpBeforeMatTop} once the compressed system is solved.  At this point, conversion matrix methods from Sec.~\ref{sec:cm-multiport} may be applied to the compressed system and combined with a loading matrix $\C{R}$ representing time-varying loads on the ports associated with the loaded currents $\M{I}_\T{l}$, leading to
\begin{equation}
    \left( \mathbf{\hat{\check{Z}}}_\T{l} + \C{R}_\T{l} \right) \C{I}_\T{l} = \mathbf{\hat{\check{V}}}_\T{l} \mathrm{.}
    \label{eq:CmpBeforeCptCM}
\end{equation}

The compressed conversion matrix system is now of dimension $N_\T{l} N_\T{f}$, leading to considerably lower inversion cost than the uncompressed $N N_\T{f}$-dimensional system.  This is particularly true when a high-dimensional LTI system has a comparatively small number of time-varying elements, i.e., $N_\T{l}\ll N$. Compression of this form in the extreme case of a single load amounts to collapsing the LTI portion of the system into a one-port impedance, equivalent to the approach taken in \cite{huang1993analysis}.  

\subsection{Computational Cost Analysis}

Two stages determine the total cost of using the previously described compression technique: construction of the compressed system matrix and its inversion. In constructing the system matrix $\mathbf{\hat{\check{Z}}}_\T{l}$, compression at each harmonic is carried out via \eqref{eq:CmpBefore}, resulting in a total cost of $\mathcal{O}(2N_\T{f}N_\T{l}N_\T{u}^2 + N_\T{f}N_\T{u}^3)$. Once constructed, the system matrix $\mathbf{\hat{\check{Z}}}_\T{l}$ has a dimension of $N_\T{l} N_\T{f}$ resulting in a na\"ive inversion cost of $\mathcal{O}(N_\T{l}^3 N_\T{f}^3)$.  Clearly, the total cost depends on the relative numbers of loaded ports and frequencies.

\begin{figure}
    \centering
    \includegraphics[width=3.5in,clip,trim=0.05in 0in 0in 0in]{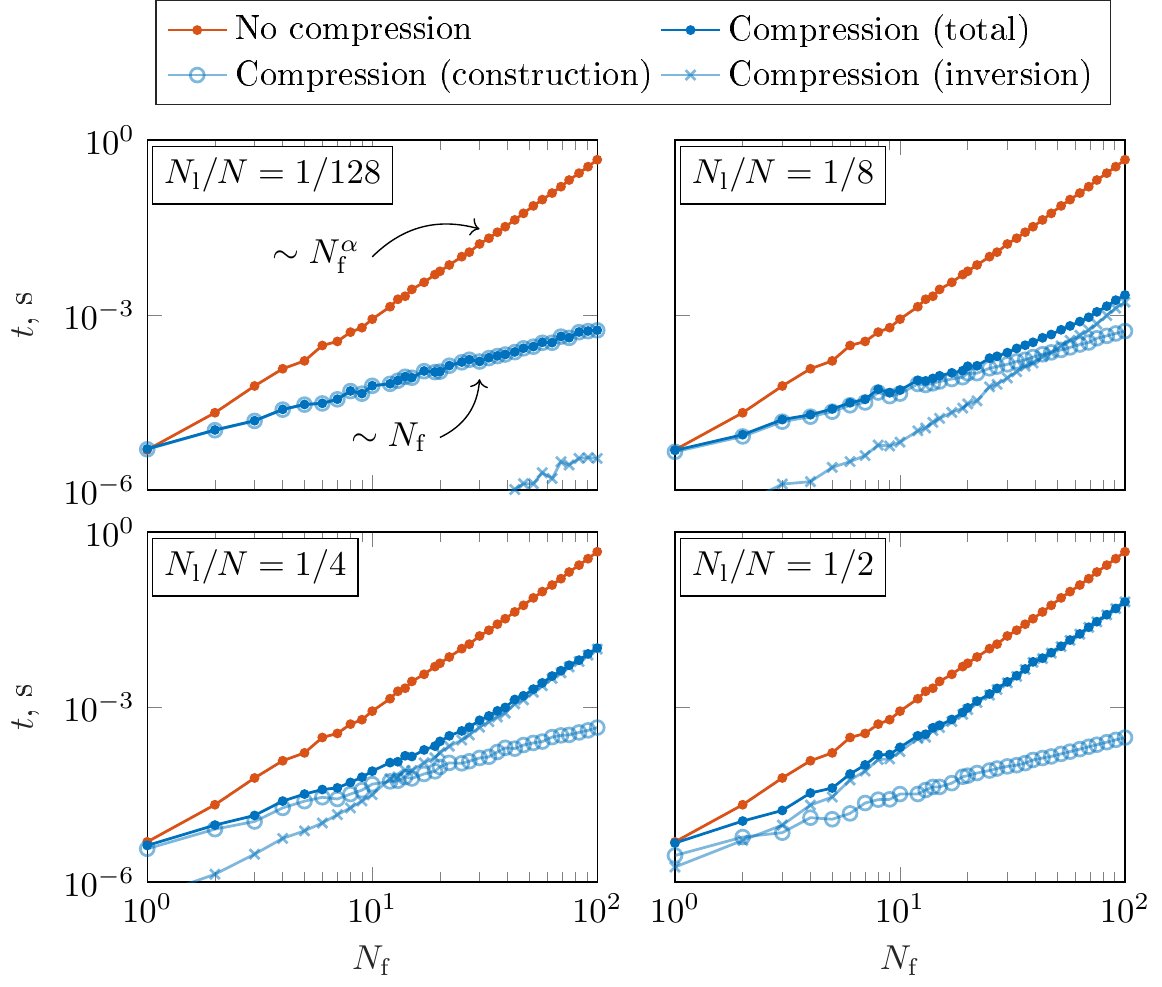}
    \caption{Conversion matrix solution times with and without compression as a function of the number of frequencies $N_f$ for varying numbers of loaded ports $N_\T{l}$.  Note that in this experiment we observe inversion costs scaling approximately as $\mathcal{O}(N^\alpha)$, with $\alpha \approx 2.75$.}
    \label{fig:CompressTime}
\end{figure}

To examine computational speedup afforded by this method of compression in a variety of scenarios, we compute the time\footnote{Inversion is carried out by the MATLAB function \texttt{inv} with $N = 128$.   In all timing experiments, random dense matrices are used and median times are recorded based on 100 sequential trials.} taken to invert a matrix of dimension $N_\T{f}N$ with no compression applied.  Additionally, we solve the same system using compression by constructing and inverting the compressed system matrix $\mathbf{\hat{\check{Z}}}_\T{l}$ for several ratios of loaded ports $N_\T{l}/N$. Measured times from both methods are shown in Fig.~\ref{fig:CompressTime}.

In all cases studied, compression leads to significantly reduced computational cost, though the magnitude of this reduction depends on the relative time spent on construction versus inversion.  The computational cost of inverting the system matrix $\mathbf{\hat{\check{Z}}}_\T{l}$ is dominant when the number of frequency points is relatively high, while construction costs dominate when the number of frequency points is low.  This relationship is modulated by the relative number of loaded ports, as seen in the moving intersection of inversion and construction costs across all panels of Fig.~\ref{fig:CompressTime}.

\begin{figure}
    \centering
    \includegraphics[width=3.25in]{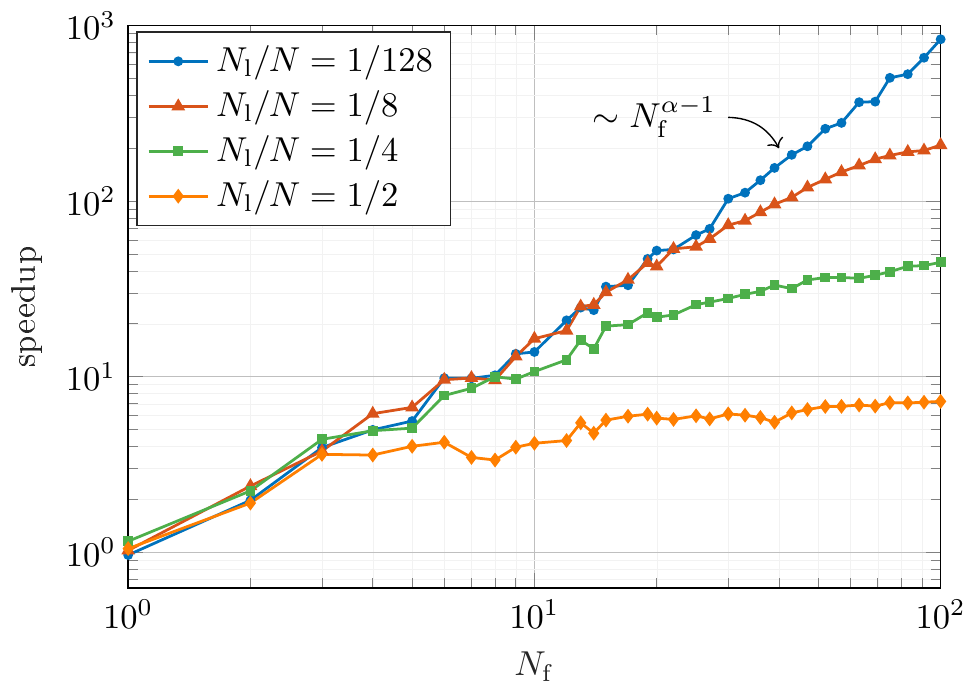}
    \caption{Speed up afforded by compression, as calculated by the quotient of total solution times recorded in Fig.~\ref{fig:CompressTime}.}
    \label{fig:CompressSpeedup}
\end{figure}

In the case of dominant construction costs, compression yields a speedup proportional to $N_\T{f}^{\alpha-1}$, where $\alpha = 3$ corresponds to nominal na\"ive inversion complexity, see footnote~1.  Conversely, when inversion costs dominate the speedup is independent of the number of frequencies and on the order of  $(N/ N_\T{l})^3$.  Both of these trends are visible in Fig.~\ref{fig:CompressSpeedup}, where the speedup (as defined by the quotient of the uncompressed and compressed computational times) is plotted using the data presented in Fig.~\ref{fig:CompressTime}.

\section{Examples}\label{sec:examples}
In this section, we present a selection of example problems solved using the hybridized conversion matrix method of moments (CMMoM) technique.  Like conventional method of moments, the proposed method is capable of modeling a broad range of practical antenna and scattering problems.  The examples included here highlight the method's application to scattering, transmission, and reception using structures with both lumped and distributed time-varying loads. 

\subsection{Scattering from a singly-loaded bowtie dipole}
\begin{figure}
    \centering
    \includegraphics[width=\linewidth]{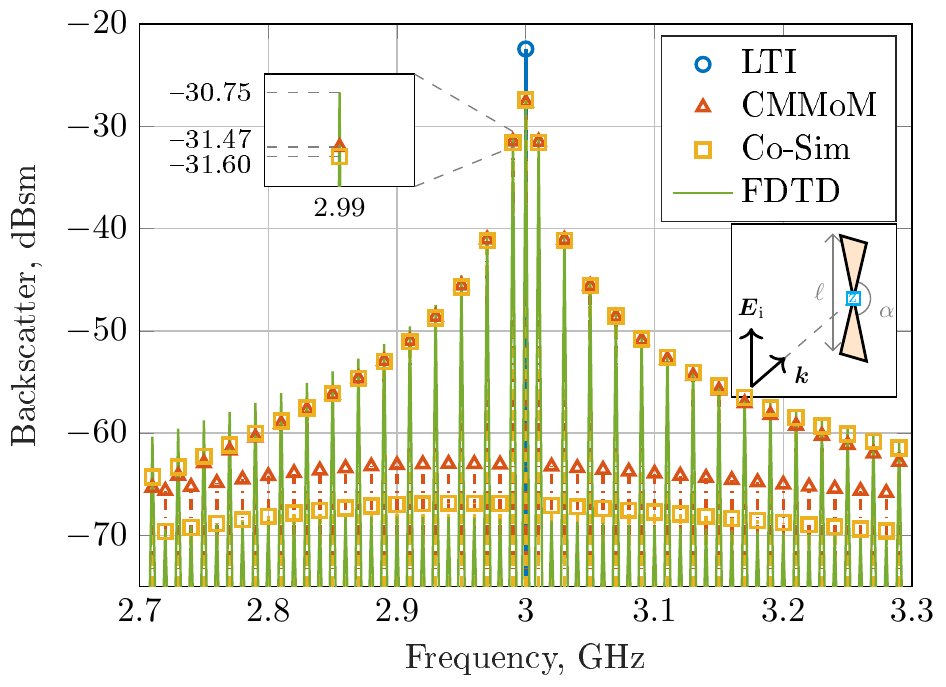}
    \caption{Cross-frequency backscatter spectra from CMMoM, transient circuit co-simulation, FDTD, and measurement, compared to backscatter from LTI bowtie.  Inset schematic.}
    \label{fig:BTSpectra}
\end{figure}

We begin with the analysis of a bowtie antenna loaded at its center by a time-varying switch, as shown in the inset of Fig.~\ref{fig:BTSpectra}.  The bowtie length $\ell$ is $36\,\T{mm}$ and the angle $\alpha$ between the two arms is $155^{\circ}$. The central switch is modeled by
\begin{equation}
    R_\T{L}(t) = 
    \begin{cases}
    r_0 & t_\T{L}/2 \leq t - nt_\T{L}< t_\T{L} \\
    0 & \T{else}
    \end{cases}
\end{equation}
with an off-resistance $r_0$ of 10 k$\Omega$ and switching frequency $f_\T{L}=1/t_\T{L}$ of $10\,\T{MHz}$.

The excitation is an incident plane wave from broadside at $3\,\T{GHz}$ co-polarized with the long dimension of the bowtie and is 
defined as
\begin{equation}
    \V{e}^\T{inc}(t,\V{r}) = \u{z}E_0 \sin{\omega_\T{inc} t}, \quad \forall~\V{r} = \u{y}y+\u{z}z
\end{equation}
in the plane of the bowtie antenna ($x = 0$) where $t$ is the same time variable shared by the load.  Fig.~\ref{fig:BTSpectra} shows the monostatic backscatter produced by this system. Note that while the excitation in this example is monochromatic, the scattered fields consist of multiple harmonics due to the time-varying load.  Similar to polarization-specific scattering analysis, here we adopt the following notation for multi-harmonic backscatter,
\begin{equation}
\BS(\omega_\T{inc},\omega_\T{obs}) = \lim_{r\rightarrow\infty}4\pi r^2 \frac{|\boldsymbol{E}^\T{sc}(\omega_\T{obs})|^2}{|\boldsymbol{E}^\T{inc}(\omega_\T{inc})|^2}
\end{equation}
where $\V{E}^\T{inc}$ and $\V{E}^\T{sc}$ are incident and backscattered fields, $r$ is a distance from the scattering object, and $\omega_\T{inc}$ and $\omega_\T{obs}$ are the incident and observation angular frequencies, respectively.

A CMMoM model of the bowtie example structure was constructed with 170~triangles, 220~RWG basis functions, and 201~harmonic frequencies. 
Fig. \ref{fig:BTSpectra} shows the agreement between CMMoM, a commercial FDTD code \cite{xfdtd}, and transient circuit co-simulation \cite{ads}. CMMoM results from a static bowtie with no time-varying load are also included for comparison. The CMMoM, FDTD, and circuit co-simulation results of the time-varying bowtie model agree within 0.3 dB at the zeroth harmonic (incident frequency) and 0.9 dB at the first-order harmonics.  There is larger relative (dB) error in the higher order harmonics, though the linear magnitudes of these differences are relatively small due to the much smaller absolute magnitude of these higher order harmonics. 

We observe that the backscatter spectrum contains primarily odd-numbered harmonics of the $500\,\T{MHz}$ square wave switching waveform, which is to be expected since the Fourier series of a square wave contains only odd numbered harmonics. It should be noted that physical systems with linear time-varying loads contain only intermodulation frequencies of the excitation signal and time-varying loading waveform. By definition, CMMoM produces output only at these discrete harmonic frequencies, which are known a priori. Time domain methods, on the other hand, can produce additional, spurious spectral content due to transient and windowing effects.

\begin{figure}
    \centering
    \includegraphics[width=\linewidth]{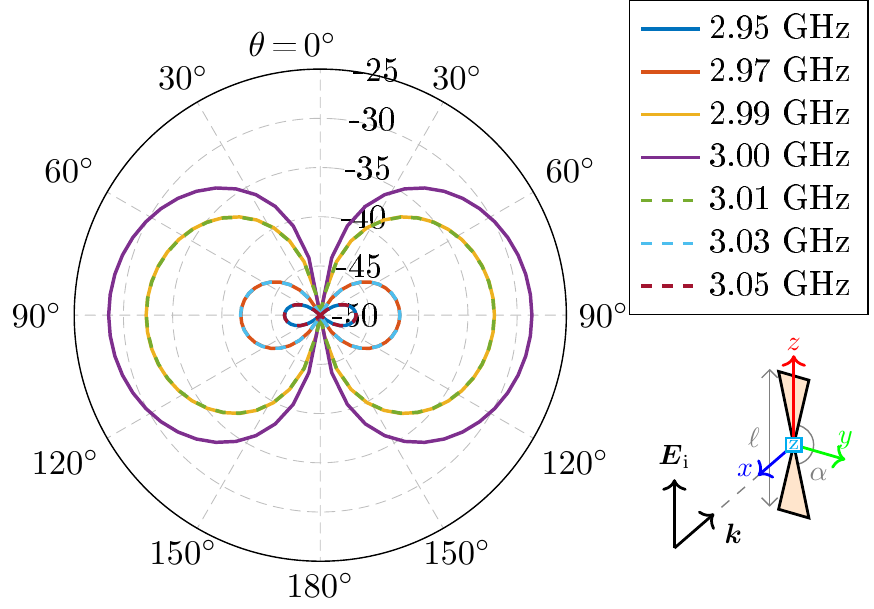}
    \caption{Normalized backscattered power (dBsm) of the bowtie example from CM-MoM vs. declination angle $\theta$ in the $xz$ plane for the frequencies of greatest returned power.}
    \label{fig:BTPattern}
\end{figure}

Figure~\ref{fig:BTPattern} shows the normalized backscattered 
power as a function of declination angle $\theta$ due to an excitation field incident from $\theta=90^{\circ}$. The pattern of the reflected field for each harmonic has the shape of a center-fed dipole, with nulls in endfire directions and peaks at broadside. The magnitude of each harmonic pattern at broadside corresponds to the peak values in Fig. \ref{fig:BTSpectra} and the values listed in Table \ref{tab:my_label}.

\begin{table}[]
    \centering
    \caption{Comparison of backscatter (dBsm) data from Fig.~\ref{fig:BTSpectra}}
    \begin{tabular}{c|c|c|c|c|c|c}
     & Static & \multicolumn{5}{c}{Time-varying}\\\hline
         & $k = 0$ & $-2$ & $-1$ & $0$ & $1$ & $2$\\\hline
        Co-sim & -21.5 & -66.9 & -31.6 & -27.4 & -31.6 & -67.2 \\
        CMMOM & -21.5 & -63.0 & -31.5 & -27.7 & -31.5 & -63.3\\
        FDTD & -21.1 & -68.4 & -30.7 & -27.5 & -30.8 & -68.6\\
    \end{tabular}
    \label{tab:my_label}
\end{table}

\subsection{Harmonic generation in a multiply-loaded transmit system}

\begin{figure}
    \centering
    \includegraphics[width=\linewidth]{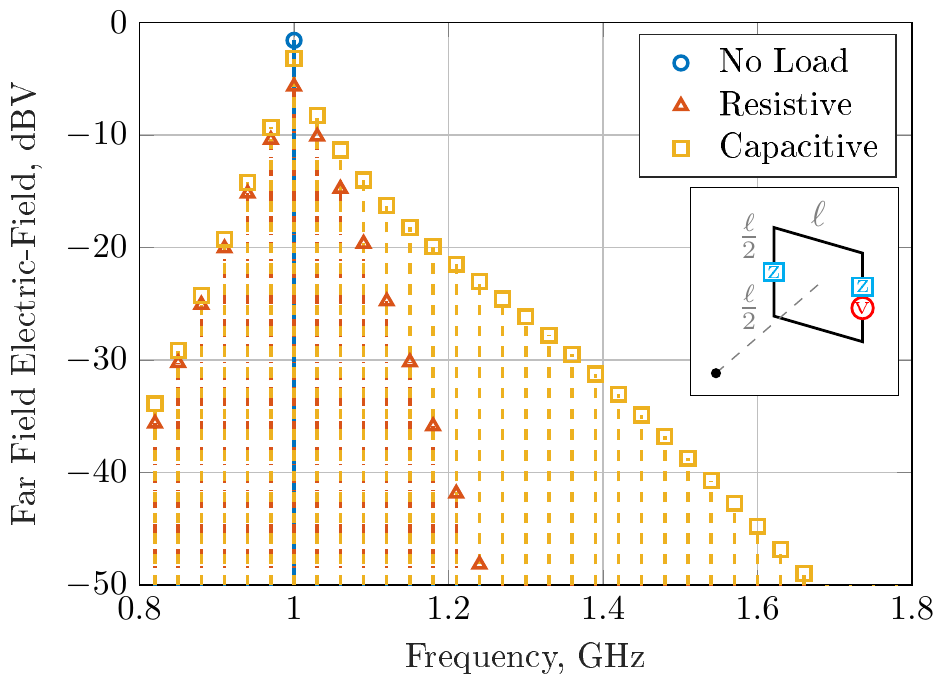}
    \caption{CMMoM calculation of broadside radiated electric fields from a transmitting square loop loaded with two time-varying resistive or capacitive loads.  Load (blue) and voltage source (red) locations are shown on the inset schematic.}
    \label{fig:LPHarmPow}
\end{figure}

As a second example, we consider a square wire loop antenna loaded on opposite sides by two sinusoidally-varying time-varying loads, selected either as time-varying resistors or time-varying capacitors. The loop has a side length $\ell$ of 82.8~mm and a radius of 1~mm. The time-varying loads are defined by
\begin{equation}
    R_\T{L}(t) = R_0\left(1+\gamma\cos\omega_L t\right),
\end{equation}
and
\begin{equation}
    C_\T{L}(t) = C_0\left(1+\gamma\cos\omega_L t\right),
\end{equation}
where
\begin{equation}
    R_0 = 150 \, \Omega \mathrm{,} \quad C_0 = 5 \, \mathrm{pF},
\end{equation}
and the frequency of the loads is set to $f_L=\omega_L/(2\pi)=30\, \mathrm{MHz}$. The modulation coefficient $\gamma$ prevents the resistance and capacitance from reaching zero, which would lead to divergent Fourier representations of the conductance and elastance. In this example, the modulation coefficient $\gamma$ is set to 0.95.  The excitation is a voltage gap feed at $1\,\T{GHz}$ located next to one of the loads and defined as
\begin{equation}
    v^\T{inc}(t) = V_0 \cos{\omega_\T{inc} t}
\end{equation}
where $V_0=1\ \T{V}$ and $t$ is the same time variable shared by the loads. The voltage source location, as well as the locations of the resistive and capacitive loads, are shown in Fig.~\ref{fig:LPHarmPow}.

The CMMoM model of the square loop is constructed with 66~rooftop basis functions with 191~harmonic frequencies. The radiated electric field in the direction normal to the loop for a $1\,\T{GHz}$ excitation is shown in Fig.~\ref{fig:LPHarmPow}. Similar to the bowtie example, the excitation frequency is modulated by the load frequency to produce harmonics, but in this case both even and odd harmonics are prominent because of the sinusoidal load waveform. The capacitive loads show more radiated power compared to the resistive loads, which are lossy by nature.

\subsection{Scattering due to distributed surface resistance}

As an example of computations involving distributed time-varying material parameters, we consider a rectangular plate with time-varying surface resistivity $R_\T{s}(\V{r},t)$.  For simplicity, we set this resistance to be isotropic and spatially homogeneous and assign the following time variation
\begin{equation}
    R_\T{s}(t) = R_0\left(1+\gamma\cos\alpha\omega_\T{c} t\right).
\end{equation}
The plate has an aspect ratio of $2:1$ and an electrical size of $ka = 0.5$ relative to the incident plane wave of frequency $\omega_\T{c}$.  The excitation is incident from the broadside direction and is co-polarized with the long dimension of the plate.  The plate is meshed with 198 RWG basis functions and impedance and Gram matrices were produced using AToM \cite{atom}.

\begin{figure}
    \centering
    \includegraphics[width=3.5in]{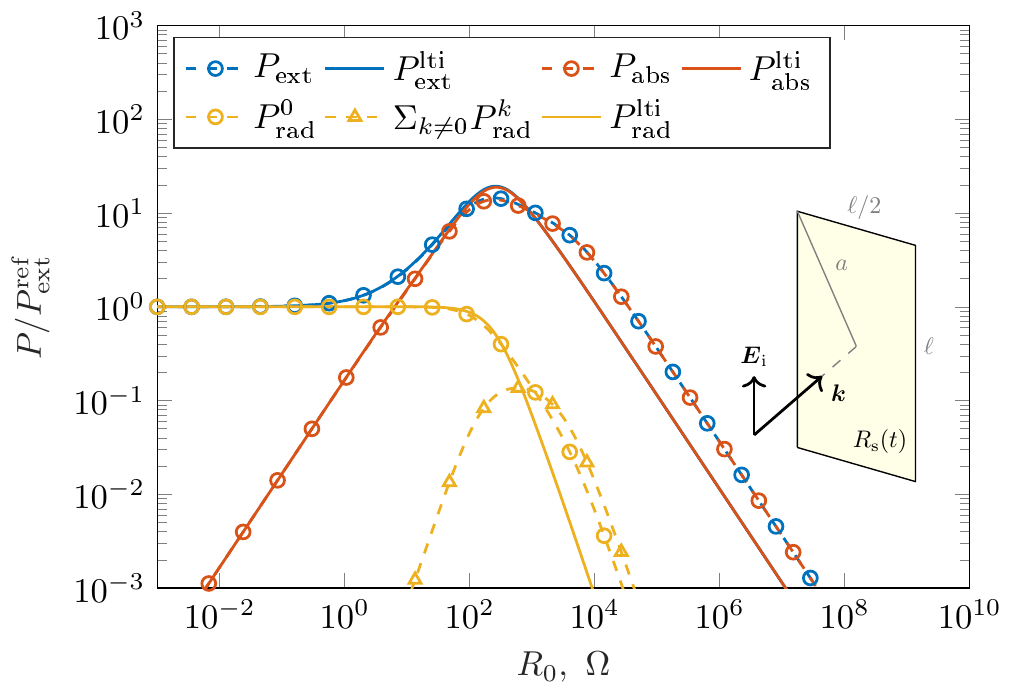}
    \caption{Scattered and extincted powers for a rectangular plate with $2:1$ aspect ratio and time-varying surface resistance $R_\T{s}(t) = R_0(1+\gamma\cos\alpha\omega_\T{c} t)$  with $\gamma = 0.95$ and $\alpha = 0.1$ illuminated at broadside by a plane wave of frequency $\omega_\T{c}$ polarized along the long dimension of the plate.  The plate has an electrical size of $ka = 0.5$ at the excitation frequency. Quantities obtained in the case of static surface resistance $R_\T{s}(t) = R_0$ are shown as solid lines for comparison.}
    \label{fig:st-totals}
\end{figure}

\begin{figure}
    \centering
    \includegraphics[width=3.5in]{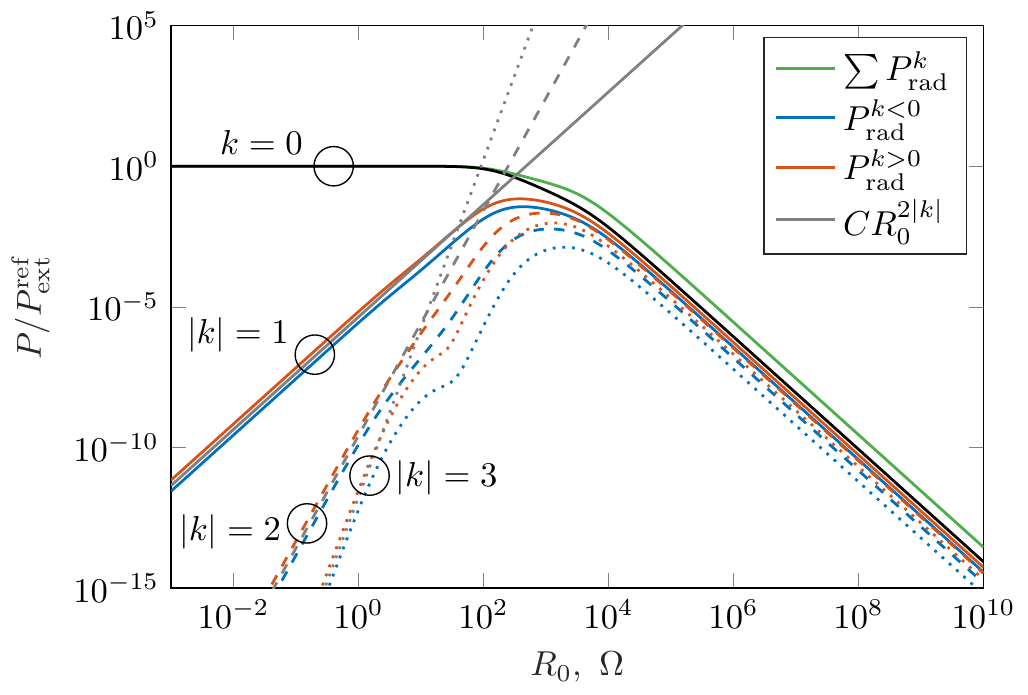}
    \caption{Breakdown of harmonic scattering from the problem in Fig.~\ref{fig:st-totals}.  Solid black line denotes scattering at the fundamental frequency.  Clusters of red and blue traces show scattering at individual frequencies.  Gray lines show $R_\T{s}^{2k}$ trends.}
    \label{fig:st-harmonics}
\end{figure}

Conversion matrix systems of the form of \eqref{eq:rs-cm} are generated for a variety of values of $R_0$ with fixed parameters $\gamma = 0.95$ and $\alpha = 0.1$ using $K = 20$ for a total of 41 harmonics\footnote{All reported quantities were well converged for $K>10$ in this particular example.}   Fig. \ref{fig:st-totals} shows the normalized total scattering, extinction, and absorption for this structure at all harmonic frequencies as a function of the time-varying surface resistance magnitude $R_0$.  For comparison, we also compute and plot all quantities for the case of a static surface resistivity $R_\T{s}(t) = R_0$.  The reference extinction power $P_\T{ext}^\T{ref}$ used for normalization is that produced by a static PEC system with $R_0 = 0~\Omega$.  Interestingly, we observe in this example that scattered harmonic generation is maximized roughly near values of $R_0$ that maximize absorption in the LTI case. In this regime, the net scattered power over all harmonics $\sum_{k\neq 0}P_\T{rad}^k$ outweighs scattered power in the fundamental frequency $P_\T{rad}^0$ by approximately one order of magnitude. A breakdown of how individual harmonic frequencies contribute to the total scattered power is shown in Fig.~\ref{fig:st-harmonics}, where it is clear that in the small loading ($R_0 \ll 1$) regime harmonic scattered powers grow as even powers of the parameter $R_0$.  Additionally, we see that in the large loading regime ($R_0 \gg 1$) the scattered powers from many individual harmonics are comparable to the power scattered at the fundamental excitation frequency.

\begin{figure}
    \centering
    \includegraphics[width=3.5in]{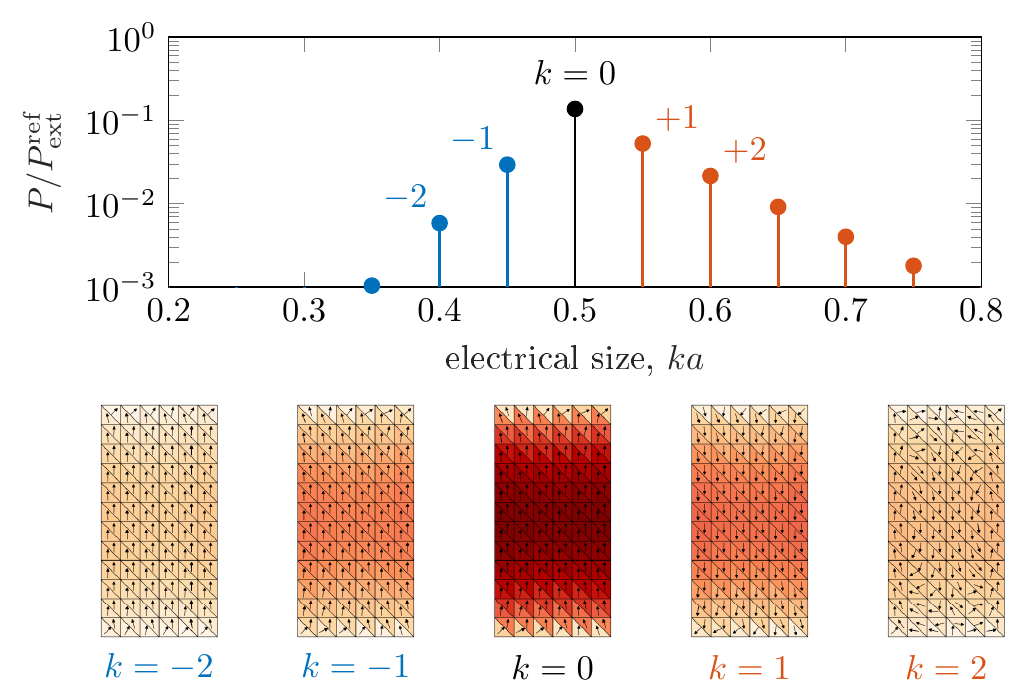}
    \caption{Harmonic scattering at individual frequencies from Fig.~\ref{fig:st-harmonics} in the case of surface resistance magnitude $R_0 = 1000~\Omega$.  Current distributions at selected harmonics are shown below, with the underlying colormap depicting the current magnitude normalized to its maximum value at the fundamental ($k=0$) frequency.}
    \label{fig:st-slice}
\end{figure}

In Fig.~\ref{fig:st-slice}, we examine the single case of $R_0 = 1000~\Omega$ and plot the scattered power and current distribution over a range of harmonic frequencies centered about the fundamental electrical size $ka = 0.5$.  The scattered powers correspond roughly to the square relative current magnitudes as all plotted harmonic currents have roughly the same distribution, with slight edge confinement observed in the higher order cases.  We note that, at all plotted harmonics, the structure is electrically small, justifying the somewhat consistent current distributions at each frequency.  In contrast to previous examples, in this calculation no cost-reducing compression may be applied as the entire system is loaded with a time-varying surface resistivity.  This leads to very large computational cost for even modest mesh densities, cf Sec.~\ref{sec:computational}.  Studying the structure at higher frequencies where harmonic currents may vary significantly in their distribution naturally requires an increased mesh density, leading to further increased computational cost.

\section{Conclusion}\label{sec:conclusion}
In this paper we present a hybridized conversion matrix-MoM (CMMoM) technique capable of modeling periodically time-varying linear loads on arbitrarily shaped structures.  We formulate the method for both lumped time-varying loads and distributed space-time modulated materials.  Numerical results from several examples demonstrate the flexibility of the proposed method and verify its accuracy against general purpose time-domain solvers.

The hybridized CMMoM method allows flexible frequency-domain analysis of a wide class of structures, but is not without limitations. First, while this method can be applied to small-signal analysis of nonlinear loads operated under locally linear conditions, it cannot model large-signal nonlinear effects. Second, very large distributed time-varying structures with large numbers of harmonics quickly lead to systems of equations requiring enormous computational effort to solve. Finally, while the study of matrix operators generated for LTI MoM structures can be informative and physically significant, CMMoM matrices lack many inherent symmetry properties and the physical interpretation of CMMoM matrix properties is less immediately clear. 

Despite these limitations, the method has a variety of uses in the modeling of electromagnetic problems ranging from direct antenna modulation to spatiotemporally modulated materials. With high opportunity for pixelized partial matrix reuse, we expect it to support development of new automated design methods for non-LTI electromagnetic structures. Its impedance-based formulation may also admit new theoretical analyses, e.g., the derivation of physical bounds, leading to the improved understanding of time-varying electromagnetic systems.

\section*{Acknowledgments}
This work was partially funded by U.S. government contract 2019-19012300001S. The authors would like to thank Jay McDaniel, Rachel Jarvis, Kyle Kanaly, and Clayton Blosser of the University of Oklahoma Advanced Radar Research Center for their assistance.

\bibliographystyle{ieeetr}
\bibliography{TVrefs, kurt-bonus-refs}

\begin{thebibliography}{10}

\bibitem{bahr1977use}
A.~Bahr, ``On the use of active coupling networks with electrically small
  receiving antennas,'' {\em {IEEE} Trans. Antennas Propag.}, vol.~25, no.~6,
  pp.~841--845, 1977.

\bibitem{nordholt1980new}
E.~Nordholt and D.~Van~Willigen, ``A new approach to active antenna design,''
  {\em {IEEE} Trans. Antennas Propag.}, vol.~28, no.~6, pp.~904--910, 1980.

\bibitem{zhu2012}
N.~{Zhu} and R.~W. {Ziolkowski}, ``Broad-bandwidth, electrically small antenna
  augmented with an internal non-{F}oster element,'' {\em {IEEE} Antennas
  Wireless Propag. Lett.}, vol.~11, pp.~1116--1120, 2012.

\bibitem{daly2018tuning}
E.~{Daly} and M.~{Daly}, ``The effect of phase continuity on synchronous
  antenna tuning,'' in {\em Proc.~Intl.~Symp.~Antennas Propag.}, pp.~517--518,
  2018.

\bibitem{schab2019pulse}
K.~Schab, D.~Huang, and J.~J. Adams, ``Pulse characteristics of a direct
  antenna modulation transmitter,'' {\em IEEE Access}, 2019.

\bibitem{santos2020}
J.~P. {Dytioco Santos}, F.~{Fereidoony}, M.~{Hedayati}, and Y.~E. {Wang},
  ``High efficiency bandwidth {VHF} electrically small antennas through direct
  antenna modulation,'' {\em {IEEE} Trans. Microw. Theory Techn.}, vol.~68,
  no.~12, pp.~5029--5041, 2020.

\bibitem{hopf1981fast}
J.~Hopf and H.~Lindenmeier, ``Fast tunable active receiving antennas,'' {\em
  Radio Science}, vol.~16, no.~6, pp.~1143--1147, 1981.

\bibitem{wang2007}
X.~{Wang}, L.~P.~B. {Katehi}, and D.~{Peroulis}, ``Time-varying matching
  networks for signal-centric systems,'' {\em {IEEE} Trans. Microw. Theory
  Techn.}, vol.~55, no.~12, pp.~2599--2613, 2007.

\bibitem{loghmannia2019parametric}
P.~{Loghmannia} and M.~{Manteghi}, ``An active cavity-backed slot antenna based
  on a parametric amplifier,'' {\em {IEEE} Trans. Antennas Propag.}, vol.~67,
  no.~10, pp.~6325--6333, 2019.

\bibitem{slevin2020}
E.~{Slevin}, P.~{Singletary}, K.~{Whitmore}, B.~{Gurses}, N.~{Opalinski},
  L.~{Thompson}, M.~B. {Cohen}, and M.~{Golkowski}, ``Broadband {VLF/LF}
  transmission from an electrically-small structure via time-varying antenna
  properties,'' in {\em IEEE APSURSI}, pp.~1663--1664, 2020.

\bibitem{singletary2021}
P.~J. {Singletary} and M.~B. {Cohen}, ``Using a high-speed plasma as a
  conducting channel to enable a novel antenna approach,'' {\em IEEE Plasma
  Sci.}, pp.~1--11, 2021.

\bibitem{koutserimpas2018electromagnetic}
T.~T. Koutserimpas and R.~Fleury, ``Electromagnetic waves in a time periodic
  medium with step-varying refractive index,'' {\em IEEE Trans.~Antennas
  Propag.}, vol.~66, no.~10, pp.~5300--5307, 2018.

\bibitem{ramaccia2018nonreciprocity}
D.~Ramaccia, D.~L. Sounas, A.~Al{\`u}, F.~Bilotti, and A.~Toscano,
  ``Nonreciprocity in antenna radiation induced by space-time varying
  metamaterial cloaks,'' {\em IEEE Antennas Wireless Propag. Lett.}, vol.~17,
  no.~11, pp.~1968--1972, 2018.

\bibitem{caloz2019spacetime_pt1}
C.~Caloz and Z.-L. Deck-L{\'e}ger, ``Spacetime metamaterials—part i: General
  concepts,'' {\em IEEE Transactions on Antennas and Propagation}, vol.~68,
  no.~3, pp.~1569--1582, 2019.

\bibitem{caloz2019spacetime_pt2}
C.~Caloz and Z.-L. Deck-L{\'e}ger, ``Spacetime metamaterials—part ii: Theory
  and applications,'' {\em IEEE Transactions on Antennas and Propagation},
  vol.~68, no.~3, pp.~1583--1598, 2019.

\bibitem{shcherbakov2019}
M.~R. Shcherbakov, P.~Shafirin, and G.~Shvets, ``Overcoming the
  efficiency-bandwidth tradeoff for optical harmonics generation using
  nonlinear time-variant resonators,'' {\em Phys. Rev. A}, vol.~100, p.~063847,
  Dec 2019.

\bibitem{chamanara2019}
N.~{Chamanara}, Y.~{Vahabzadeh}, and C.~{Caloz}, ``Simultaneous control of the
  spatial and temporal spectra of light with space-time varying metasurfaces,''
  {\em IEEE Transactions on Antennas and Propagation}, vol.~67, no.~4,
  pp.~2430--2441, 2019.

\bibitem{wu2019serrodyne}
Z.~Wu and A.~Grbic, ``Serrodyne frequency translation using time-modulated
  metasurfaces,'' {\em IEEE Trans.~Antennas Propag.}, vol.~68, no.~3,
  pp.~1599--1606, 2019.

\bibitem{taravati2020space}
S.~Taravati and A.~A. Kishk, ``Space-time modulation: Principles and
  applications,'' {\em IEEE Microw.~Mag.}, vol.~21, no.~4, pp.~30--56, 2020.

\bibitem{landt1983time}
J.~Landt, E.~Miller, and F.~Deadrick, ``Time domain modeling of nonlinear
  loads,'' {\em {IEEE} Trans. Antennas Propag.}, vol.~31, no.~1, pp.~121--126,
  1983.

\bibitem{du2019simulation}
Z.-X. Du, A.~Li, X.~Y. Zhang, and D.~F. Sievenpiper, ``A simulation technique
  for radiation properties of time-varying media based on frequency-domain
  solvers,'' {\em IEEE Access}, vol.~7, pp.~112375--112383, 2019.

\bibitem{garbacz1971generalized}
R.~Garbacz and R.~Turpin, ``A generalized expansion for radiated and scattered
  fields,'' {\em IEEE Trans.~Antennas Propag.}, vol.~19, no.~3, pp.~348--358,
  1971.

\bibitem{harrington1971theory}
R.~Harrington and J.~Mautz, ``Theory of characteristic modes for conducting
  bodies,'' {\em IEEE Trans.~Antennas Propag.}, vol.~19, no.~5, pp.~622--628,
  1971.

\bibitem{rahmat1999electromagnetic}
Y.~Rahmat-Samii and E.~Michielssen, ``Electromagnetic optimization by genetic
  algorithms,'' {\em Microw.~J.}, vol.~42, no.~11, pp.~232--232, 1999.

\bibitem{ethier2014antenna}
J.~L. Ethier and D.~A. McNamara, ``Antenna shape synthesis without prior
  specification of the feedpoint locations,'' {\em IEEE Trans.~Antennas
  Propag.}, vol.~62, no.~10, pp.~4919--4934, 2014.

\bibitem{capek2019shape}
M.~Capek, L.~Jelinek, and M.~Gustafsson, ``Shape synthesis based on topology
  sensitivity,'' {\em IEEE Trans.~Antennas Propag.}, vol.~67, no.~6,
  pp.~3889--3901, 2019.

\bibitem{gustafsson2016antenna}
M.~Gustafsson, D.~Tayli, C.~Ehrenborg, M.~Cismasu, and S.~Nordebo, ``Antenna
  current optimization using {MATLAB} and {CVX},'' {\em FERMAT}, vol.~15,
  no.~5, pp.~1--29, 2016.

\bibitem{gustafsson2019tradeoff}
M.~Gustafsson, M.~Capek, and K.~Schab, ``Tradeoff between antenna efficiency
  and {Q}-factor,'' {\em IEEE Trans.~Antennas Propag.}, vol.~67, no.~4,
  pp.~2482--2493, 2019.

\bibitem{huang1993analysis}
C.-C. Huang and T.-H. Chu, ``Analysis of wire scatterers with nonlinear or
  time-harmonic loads in the frequency domain,'' {\em {IEEE} Trans. Antennas
  Propag.}, vol.~41, no.~1, pp.~25--30, 1993.

\bibitem{jayathurathnage2021timevarying}
P.~Jayathurathnage, F.~Liu, M.~S. Mirmoosa, X.~Wang, R.~Fleury, and S.~A.
  Tretyakov, ``Time-varying components for enhancing wireless transfer of power
  and information.'' arXiv, 2021.

\bibitem{epp1992}
L.~W. {Epp}, C.~H. {Chan}, and R.~{Mittra}, ``Periodic structures with
  time-varying loads,'' {\em {IEEE} Trans. Antennas Propag.}, vol.~40, no.~3,
  pp.~251--256, 1992.

\bibitem{salary2018time}
M.~M. Salary, S.~Jafar-Zanjani, and H.~Mosallaei, ``Time-varying metamaterials
  based on graphene-wrapped microwires: Modeling and potential applications,''
  {\em Phys.~Rev.~B}, vol.~97, no.~11, p.~115421, 2018.

\bibitem{palmer2019investigation}
A.~Palmer, ``Investigation of a generalized frequency domain method for
  modeling time-varying loads on antennas,'' Master's thesis, University of
  Oklahoma, 2019.

\bibitem{maas2003nonlinear}
S.~A. Maas, {\em Nonlinear microwave and RF circuits}.
\newblock Artech House, 2003.

\bibitem{jiang2006conversion}
C.~Jiang, T.~K. Johanson, and V.~Krozer, ``Conversion matrix analysis of gaas
  hemt active gilbert cell mixers,'' in {\em INMMIC}, pp.~94--97, IEEE, 2006.

\bibitem{harrington1993field}
R.~F. Harrington, {\em Field computation by moment methods}.
\newblock Wiley-IEEE Press, 1993.

\bibitem{rao1982electromagnetic}
S.~Rao, D.~Wilton, and A.~Glisson, ``Electromagnetic scattering by surfaces of
  arbitrary shape,'' {\em {IEEE} Trans. Antennas Propag.}, vol.~30, no.~3,
  pp.~409--418, 1982.

\bibitem{yeh1967theory}
Y.~Yeh and K.~Mei, ``Theory of conical equiangular-spiral antennas--{P}art
  {I}--{N}umerical technique,'' {\em IEEE Trans.~Antennas Propag.}, vol.~15,
  no.~5, pp.~634--639, 1967.

\bibitem{jiao1999fast}
D.~Jiao and J.-M. Jin, ``Fast frequency-sweep analysis of {RF} coils for
  {MRI},'' {\em IEEE Trans.~Biomed.~Eng.}, vol.~46, no.~11, pp.~1387--1390,
  1999.

\bibitem{lo2011finite}
Y.~H. Lo, S.~He, L.~Jiang, and W.~C. Chew, ``Finite-width gap excitation and
  impedance models,'' in {\em Proc.~Intl.~Symp.~Antennas Propag.},
  pp.~1297--1300, IEEE, 2011.

\bibitem{Jackson:100964}
J.~D. Jackson, {\em {Classical electrodynamics; 2nd ed.}}
\newblock New York, NY: Wiley, 1975.

\bibitem{gustafsson2020upper}
M.~Gustafsson, K.~Schab, L.~Jelinek, and M.~Capek, ``Upper bounds on absorption
  and scattering,'' {\em New J.~Phys.}, 2020.

\bibitem{jelinek2016optimal}
L.~Jelinek and M.~Capek, ``Optimal currents on arbitrarily shaped surfaces,''
  {\em IEEE Trans.~Antennas Propag.}, vol.~65, no.~1, pp.~329--341, 2016.

\bibitem{manley1956}
J.~M. {Manley} and H.~E. {Rowe}, ``Some general properties of nonlinear
  elements-{Part I}. {G}eneral energy relations,'' {\em Proc. of the IRE},
  vol.~44, no.~7, pp.~904--913, 1956.

\bibitem{xfdtd}
{XFDTD, Remcom 2020}.

\bibitem{ads}
{Advanced Design System, Keysight 2020}.

\bibitem{atom}
{A}ntenna {T}oolbox for {MATLAB} ({AToM}){,} {C}zech {T}echnical
  {U}niversity~in {P}rague, 2017.

\end{thebibliography}

\end{document}